\documentclass[12pt]{article}

\usepackage[english]{babel}
\usepackage{amssymb,latexsym, amsmath}

\begin{document}

\title
{Exotic instantons in eight dimensions}
\author
{E.K. Loginov\footnote{{\it E-mail address:} ek.loginov@mail.ru} and E.D. Loginova\footnote{{\it E-mail address:} loginova.ed@mail.ru}\\
\it Ivanovo State University, Ivanovo, 153025, Russia}
\date{}
\maketitle

\begin{abstract}
In this paper, we study the (anti-)self-duality equations $\ast F\wedge F=\pm F\wedge F$ in the eight-dimensional Euclidean space. Using properties of the Clifford algebra $Cl_{0,8}(\mathbb{R})$, we find a new solution to these equations.
\end{abstract}

\section{Introduction}

The discovery of instanton solutions to the Yang-Mills field equations in the four-dimensional Euclidean space has led to an intensive study of such theory and the search for multidimensional generalizations of the self-duality equations. In Refs.~\cite{corr83,ward84}, such equations were found and classified. These were first-order equations that satisfy the Yang-Mills field equations as a consequence of the Bianchi identity. Later, solutions to these equations were found, see Refs.~\cite{fair84,fubi85,corr85,ivan92,logi04,logi05,duna12}, and then used to construct classical solitonic solutions of the low energy effective theory of the heterotic string.
\par
Another approach to the construction of self-duality equations was proposed in Ref.~\cite{tchr80}. In this work, it was considered self-duality relations between higher-order terms of the field strength. An example of instantons satisfying such self-duality relations was obtained in Ref.~\cite{gros84}, see also Refs.~\cite{naka16,logi20}. As it turned out, these instantons play a role in smoothing out the singularity of heterotic string soliton solutions by incorporating one-loop corrections. Therefore, these exotic solutions were used to construct various string and membrane solutions, see Refs.~\cite{duff91,olsen00,mina01,pedd08,bill09,bill09a,bill21} and to study the higher dimensional quantum Hall effect, see Refs.~\cite{bern03,hase14,inou21}.
\par
In this paper, we study the (anti-)self-duality equations in the eight-dimensional Euclidean space with the flat metric. We use properties of the Clifford algebra $Cl_{0,8}(\mathbb{R})$ for to find new solutions of the equations.

\section{The self-duality equations}

In this section, we give a brief summary of Clifford algebras and related constructions. We list the features of the mathematical structure as far as they are of relevance to our work.
\par
We recall that the Clifford algebra $Cl_{0,8}(\mathbb{R})$ is a real associative algebra generated by the elements $\Gamma_1,\Gamma_2,\dots,\Gamma_8$ and defined by the relations
\begin{equation}
\Gamma_i\Gamma_j+\Gamma_j\Gamma_i=-2\delta_{ij}.
\end{equation}
Its subalgebra $Cl_{0,8}^0(\mathbb{R})$ generated by the elements $\Gamma_{ij}=(\Gamma_i\Gamma_j-\Gamma_j\Gamma_i)/2$ is called its even subalgebra. It can be shown that the even subalgebra of $Cl_{0,8}(\mathbb{R})$ is isomorphic to $Cl_{0,7}(\mathbb{R})$. The element $\Gamma_9=\Gamma_1\Gamma_2\dots \Gamma_7$ commutes with all other elements of $Cl_{0,8}^0(\mathbb{R})$, and its square $\Gamma_9\Gamma_9=1$. Therefore the pair $\Gamma^{\pm}=\frac{1}{2}(1\pm\Gamma_9)$ are a complete system of mutually orthogonal central idempotents, and hence the subalgebra decomposes into the direct sum of two ideals. It can be shown that these ideals are isomorphic to the algebra $M_{8}(\mathbb{R})$ of all real matrices of size $8\times 8$.
\par
Suppose $\phi: Cl_{0,8}^0(\mathbb{R})\to M_{8}(\mathbb{R})$, $\Gamma_{ij}\to R_{ij}$ is the homomorphism of the algebras with the kernel $\text{Ker}\,\phi=\{\alpha(1-\Gamma_9)\mid \alpha\in\mathbb{R}\}$. In turn, the homomorphism of the algebras induces the homomorphism $Spin(8)\to SO(8)$ of the group. Therefore the matrices $R_{ij}$ are generators of $SO(8)$. Now, note (see, e.g.,~\cite{kenn81}) that the identity
\begin{equation}\label{01}
\Gamma_{i_1\dots i_k}=\frac{1}{(8-k)!}\varepsilon_{i_1\dots i_8}\Gamma_9\Gamma^{i_8\dots i_{k+1}},
\end{equation}
holds in the Clifford algebra $Cl_{0,8}(\mathbb{R})$. Here $\varepsilon_{i_1\dots i_8}$ is the Levi-Civita symbol in eight dimensions and $\Gamma_{i_1\dots i_k}=\Gamma_{[i_1\dots i_k]}$, where the square bracket stands for the anti-symmetrization of indices with the weight $1/k!$. Choosing $k=4$ in (\ref{01}), we obtain the self-duality equations
\begin{equation}
R_{[mn}R_{ps]}=\frac{1}{24}\varepsilon_{mnpsijkl}R_{[ij}R_{kl]}.
\end{equation}
\par
Any totally antisymmetric eight-dimensional tensor of fourth rank can be written as the sum of the self-dual and the anti-self-dual parts $F_{mnps}=F^{+}_{mnps}+F^{-}_{mnps}$, where
\begin{equation}
F_{mnps}^{\pm}=\left(\delta^i_{[m}\delta^j_n\delta^k_p\delta^l_{s]}\pm\frac{1}{24}\varepsilon_{mnpsijkl}\right)F_{ijkl}.
\end{equation}
If we now use the identity
\begin{equation}\label{02}
\Gamma_{p}\Gamma_{s_1\dots s_k}=\Gamma_{ps_1\dots s_k}+\sum^k_{i=1}(-1)^i\delta_{ps_i}\Gamma_{ps_1\dots\hat{s}_{i}\dots s_k},
\end{equation}
then we obtain the following expression for the self-dual tensor
\begin{equation}
F_{mnps}^{+}=\frac{1}{24}\text{Tr}\,(R_{mn}R_{ps}R_{ij}R_{kl})F_{ijkl}.
\end{equation}
Thus, the tensor $F_{mnps}$ is anti-self-dual if $R_{mn}R_{ps}F_{mnps}=0$. In particular, the tensor $F_{mnps}=F_{[mn}F_{ps]}$ is anti-self-dual if $R_{mn}F_{mn}=0$. Note that the last equality is a sufficient condition for the anti-self-duality, but not necessary. Note also that previously known solutions do not satisfy that condition.

\section{Instantons in eight dimensions}

In this section, we find solutions of the anti-self-duality equations in the eight-dimensional Euclidean space. This equation is given by the formula
\begin{equation}\label{04}
F_{[mn}F_{ps]}=-\frac{1}{24}\varepsilon_{mnpsijkl}F_{[ij}F_{kl]},
\end{equation}
where the gauge field strength
\begin{equation}\label{30}
F_{mn}=\partial_mA_n-\partial_nA_m+[A_m,A_n]
\end{equation}
and the potential $A_m$ takes values in the Lie algebra $so(8)$.
\par
We choose the ansatz
\begin{equation}\label{05}
A_m=-\frac{1}{2}R_{mp}\partial_{p}\varphi,
\end{equation}
where $\varphi$ is a function of $x^2=x_nx^n$. To find the gauge field strength, we substitute the potential (\ref{05}) into (\ref{30}) and use the identity
\begin{equation}\label{06}
R_{mp}R_{ns}=R_{[mp}R_{ns]}+\delta_{mn}R_{ps}-\delta_{ms}R_{pn}-\delta_{pn}R_{ms}+\delta_{ps}R_{mn}
+\delta_{ms}\delta_{pn}-\delta_{mn}\delta_{ps},
\end{equation}
which is a consequence of (\ref{02}). As a result, we get
\begin{equation}\label{08}
F_{mn}=\frac{1}{2}[R_{ms}(\partial_n\partial_s\varphi-\partial_n\varphi\partial_s\varphi)
-R_{ns}(\partial_m\partial_s\varphi-\partial_m\varphi\partial_s\varphi)+R_{mn}(\partial_s\varphi)^2].
\end{equation}
Now we impose the anti-self-duality condition $R_{mn}F_{mn}=0$ and use the identities
\begin{equation}\label{30}
R_{mn}R_{sn}=\delta_{nn}(R_{ms}-\delta_{ms}),\quad R_{mn}R_{mn}=-\delta_{mm}\delta_{nn},
\end{equation}
which are consequences of (\ref{06}). As a result, we obtain the equation
\begin{equation}
\partial_s\partial_s\varphi+3\partial_s\varphi\partial_s\varphi=0.
\end{equation}
Since $\varphi=\varphi(x^2)$, this equation is equivalent to the ordinary differential equation
\begin{equation}
x^2\varphi''+4\varphi'+3x^2(\varphi')^2=0.
\end{equation}
where $\varphi'=\partial\varphi/\partial (x^2)$. Solving this equation, we find
\begin{equation}\label{07}
\varphi=\frac{1}{3}\ln\left(c_1+\frac{c_2}{x^6}\right).
\end{equation}
This is a solution of the anti-self-duality equations (\ref{04}).
\par
Interestingly, the ansatz (\ref{05}) is a solution to the self-duality equations as $\varphi=\ln(\lambda^2+x^2)$. To show this, we represent the matrices $R_{mn}$ in the following form
\begin{equation}
R_{mn}=\frac{1}{2}(e^t_me_n-e^t_ne_m),
\end{equation}
where $e_8$ is the unit $8\times8$ matrix, $e_m$ is an image of $\Gamma_m\in Cl_{0,7}(\mathbb{R})$ as $m\ne8$, and $e_m^t$ signifies the transposition of the matrix $e_m$. In the case, the potential
\begin{equation}
\tilde{A}_m=-\frac{R_{mn}x_n}{\lambda^2+x^2}
\end{equation}
and the gauge field strength
\begin{equation}\label{09}
\tilde{F}_{mn}=\frac{2\lambda^2R_{mn}}{(\lambda^2+x^2)^2}.
\end{equation}
This is exactly a solution of the self-duality equations that was obtained in Ref.~\cite{gros84}.
\par
Let us return again to the obtained solution of the anti-self-duality equations. If we substitute the solution (\ref{07}) into (\ref{08}), then we get the gauge field strength
\begin{equation}\label{10}
F_{mn}=\frac{2\lambda^2x^2}{(\lambda^2+x^6)^2}(4R_{mp}x_nx_p-4R_{np}x_mx_p-R_{mn}x^2),
\end{equation}
where $\lambda^2=c_2/c_1$. It is not difficult to see that the solutions (\ref{09}) and (\ref{10}) retain heir form when $x_n$ is replaced by $x_n-b_n$, where $b_n\in\mathbb{R}$. Further, the gauge transformations of $A(x)$ and $\tilde{A}(x)$ induce the transformations $R_{mn}\to U^{-1}R_{mn}U$, there $U\in Spin(8)$, which only lead to a change in the basis of $so(8)$ and therefore leaves the solutions unchanged. Consequently, the solutions (\ref{09}) and (\ref{10}) have the same number of free parameters and the same gauge group. At the same time, the potentials $A(x)$ and $\tilde{A}(x)$ are gauge nonequivalent. To show this, it suffices to note that
\begin{equation}
\text{tr}F^2_{mn}=-56^2\frac{4\lambda^4x^8}{(\lambda^2+x^6)^4}\ne\text{tr}\tilde{F}^2_{mn}.
\end{equation}
Note that the field strength (\ref{10}) is not a function of $x^2$ and therefore the found solution is not rotationally invariant. This fundamentally distinguishes it from the known solutions to the anti-auto-duality equations in eight dimensions.
\par
Thus, the formula (\ref{07}) indeed gives a new solution of the anti-self-duality equations (\ref{04}). Note also that the resulting anti-self-dual solution becomes self-dual and the self-dual solution be anti-self-dual if, instead of $\phi$, we will use the homomorphism $\phi':Cl_{0,8}^0(\mathbb{R})\to M_{8}(\mathbb{R})$ with the kernel $\text{Ker}\,\phi'=\{\alpha(1+\Gamma_9)\mid \alpha\in\mathbb{R}\}$.

\section{D7-brane effective action}

We now consider the Euclidean $D7$-brane in Type IIB string theory. On the world-volume of this $D7$-brane, there is an eight-dimensional Yang-Mills theory which is naturally realized as low-energy effective field theory. In order to see the (anti-)self-dual instanton effects of the obtained solutions, we consider the $\alpha'$ corrections to the gauge theory. The  gauge part of the effective action can be written as
\begin{equation}
S_{D}=S_2+S_4+\dots,
\end{equation}
where the first term is the quadratic Yang-Mills action in eight dimensions
\begin{equation}
S_2=\frac{1}{2g_{YM}^2}\int d^8x\,\text{tr}(F^2),
\end{equation}
while the second part is a quartic action of the form
\begin{equation}\label{31}
S_4=\frac{(4\pi\alpha')^2}{4!g_{YM}^2}\int d^8x\,\text{tr}(t_8F^4)-2\pi iC_0k.
\end{equation}
Here $t_8$ is the ten-dimensional extension of the eight-dimensional light-cone gauge ``zero-mode'' tensor, i.e.
\begin{align}
t_8F^4&=F^{MN}F_{PN}F_{MS}F^{PS}+\frac{1}{2}F^{MN}F_{PN}F^{PS}F_{MS}\notag\\
&-\frac{1}{4}F^{MN}F_{MN}F^{PS}F_{PS}-\frac{1}{8}F^{MN}F^{PS}F_{MN}F_{PS}.
\end{align}
Moreover, $k$ is the fourth Chern number
\begin{equation}\label{24}
k=\frac{1}{4!(2\pi)^4}\int\text{tr}(F\wedge F\wedge F\wedge F)
\end{equation}
and $C_0$ is a scalar field of the closed string RR sector.
\par
Following~\cite{bill09}, we will interpret the eight-dimensional instantons as the $D$-instantons, i.e. as instantons embedded in $D7$-branes. Such instantons are sources for RR 0-form $C_0$. When the (anti-)self-duality condition holds, the trace
\begin{equation}
\text{tr}(t_8F^4)=\pm\frac{1}{2}\text{tr}(F\wedge F\wedge F\wedge F),
\end{equation}
and hence the quartic action $S_4$ becomes
\begin{equation}
S_4=-2\pi i(C_0\pm\frac{i}{g_s})k.
\end{equation}
This precisely matches the action of the action of $k$ $D$-instantons. Thus, the eight-dimensional (anti-)self-dual instantons become the $D$-instantons when $S_2=0$, $S_4\ne0$, and all the $O(\alpha'^4/g_{YM}^2)$ terms vanish. The second condition is fulfilled in the zero-slope limit $\alpha'\to0$ with fixed $\alpha'^2/g_{YM}^2$.
\par
On the other side, it follows from the identities (\ref{30}) that $\text{tr}F_{mnps}^2=0$ and therefore
\begin{equation}
\frac{1}{12}\text{tr}(t_8F^4)=\frac{1}{4!\cdot 2^4}\text{tr}(\epsilon^{mnpsijkl}F_{mn}F_{ps}F_{ij}F_{kl})=0.
\end{equation}
Hence the fourth Chern number and the quartic action (\ref{31}) are equal to zero. Thus, the solution to the anti-self-duality equations found in the previous section is not the $D$-instanton embedded in the $D7$-brane.

\end{document}